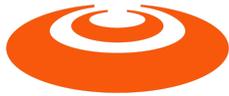

**Whitepaper**

# Education in the age of Generative AI

## Context and Recent Developments

Rafael Ferreira Mello[1], Elyda Freitas[1,2]; Filipe Dwan Pereira[1]; Luciano Cabral[1,3], Patricia Tedesco[4], Geber Ramalho[4]

[1]CESAR Innovation Center
Recife, Brazil

[2]University of Pernambuco - UPE
Caruaru, Brazil

[3]Federal Institute of Pernambuco - IFPE
Jaboatão de Guararapes, Brazil

[4]Centro de Informática - UFPE
Recife, Brazil

rflm@cesar.org.br
{elsxf,fdp,lsc6}@cesar.school
{pcart,glr}@cin.ufpe.br

## Abstract

With the emergence of generative artificial intelligence, an increasing number of individuals and organizations have begun exploring its potential to enhance productivity and improve product quality across various sectors. The field of education is no exception. However, it is vital to notice that artificial intelligence adoption in education dates back to the 1960s. In light of this historical context, this white paper serves as the inaugural piece in a four-part series that elucidates the role of AI in education. The series delves into topics such as its potential, successful applications, limitations, ethical considerations, and future trends. This initial article provides a comprehensive overview of the field, highlighting the recent developments within the generative artificial intelligence sphere.

## Introduction

Artificial Intelligence (AI) has always been portrayed in movies, such as "The Matrix," "Her," and "Ex Machina". Sometimes the AI is the villain, sometimes it is a vital solution to plot conflicts. Some films present a world where AI is seamlessly integrated into everyday life, reflecting its pervasive nature, while others offer futuristic scenarios that have, in fact, foreseen many technologies that exist today. The cinematic portrayal of AI continues to captivate audiences, showcasing its multifaceted nature and its potential implications for humanity.

In practice, the presence of AI is far more integrated into our daily lives than we may perceive. Even the simplest tasks such as searching the internet for specific data, engaging with social media platforms, seeking directions on a navigation application, or tuning into our favorite music, benefit from AI algorithms to optimize the user experience. These sophisticated systems, working silently in the background, optimize our experiences, changing how we interact with the world.

In education, it is important to note that AI's application did not begin recently. Instead, studies exploring AI in education date back to the 1960s, highlighting the long-standing interest and ongoing research in this field. In a comprehensive historical analysis of the last two decades of Artificial Intelligence in Education (AIED), Guan, Mou and Jiang [1] shed light on the development of this field. AI in education, which has gained significant attention with recent advancements like ChatGPT, traces its roots back to earlier milestones. For example, in 1964, Joseph Weizenbaum created Eliza [2], a pioneering chatbot capable of conversing with humans. Eliza is recognized as the first-ever chatbot used for education. In the late 1960s, Jaime Carbonell introduced SCHOLAR [3], an innovative program that could automatically answer questions about South American geography in natural language and assess students' responses. SCHOLAR was one of the initial models that are now widely recognized as Intelligent Tutoring Systems (ITS).

The intersection of AI and Education encompasses numerous subfields beyond AIED and ITS. Within this dynamic field, several notable areas have emerged, including Educational Data Mining (EDM), Learning Analytics (LA), and Learning at Scale (L@S), among others. These subfields contribute diverse perspectives and methodologies to the integration of AI in educational contexts. In any case, it is important to mention that AIED, as described by the International AIED Society (IAIED), "is an interdisciplinary community at the frontiers of the fields of computer science, education, and psychology" [4]. Recently these societies created the IAALDE [5], an alliance to align the vision of the different communities in order to maximize the reach of AI in Education.

The recent advancements in generative AI have significantly influenced all fields related to the AIED community. It created new opportunities for educational institutions to enhance the quality of their learning experience from different perspectives, such as assessment, feedback, and engagement, among others [6]. By adoption generative AI educational institutions can provide better instructional support for students and teachers. In this context, generative AI holds immense promise in shaping the future of education by introducing novel approaches to optimize learning experiences and facilitate educational growth. However, the negative implications need to be carefully studied before large-scale applications.

This article is the introductory piece in a comprehensive series of papers that explore the potential, successful applications, limitations, ethical considerations, and future trends of generative AI in education. In this initial article, we provide a holistic overview of the topic and main concepts and delve into critical issues that must be discussed to facilitate the effective integration of generative AI within educational institutions. By addressing these crucial aspects, we aim to pave the way for a thoughtful and efficient implementation of generative AI in education, ensuring its positive impact on teaching and learning.

# Brief History of AIED

As previously mentioned, AIED is a well-established field extensively studied since the 1960s. Before delving into the realm of generative AI, we would like to provide a timeline of significant milestones within the field. In the pursuit of showcasing the potential of utilizing generative AI, this white paper adopts the assistance of GPT4 to generate (curated by humans) a compilation of the most notable facts and developments in AIED throughout the years presented in Figure 1. By presenting this timeline, we aim to offer a comprehensive understanding of the historical progress in AIED, setting the stage for a deeper exploration of the transformative possibilities that generative AI can bring to educational settings.

In the early decades of AIED, the field largely benefited from utilizing knowledge bases and expert systems, resulting in focused and domain-specific applications such as chatbots ELIZA and SCHOLAR, as well as the initial versions of the intelligent tutoring systems. However, with the evolution of machine learning, natural language processing, and virtual reality techniques, AI's educational applications began to exhibit greater sophistication and breadth, encompassing various disciplines, methodologies, and academic tasks such as assessment, feedback, and skills development using VR and AR. More recently, from 2016 onwards, AI has become so pervasive that it is challenging to distinguish between the technique and its application, especially with the rise of sophisticated AI-driven adaptive learning platforms.

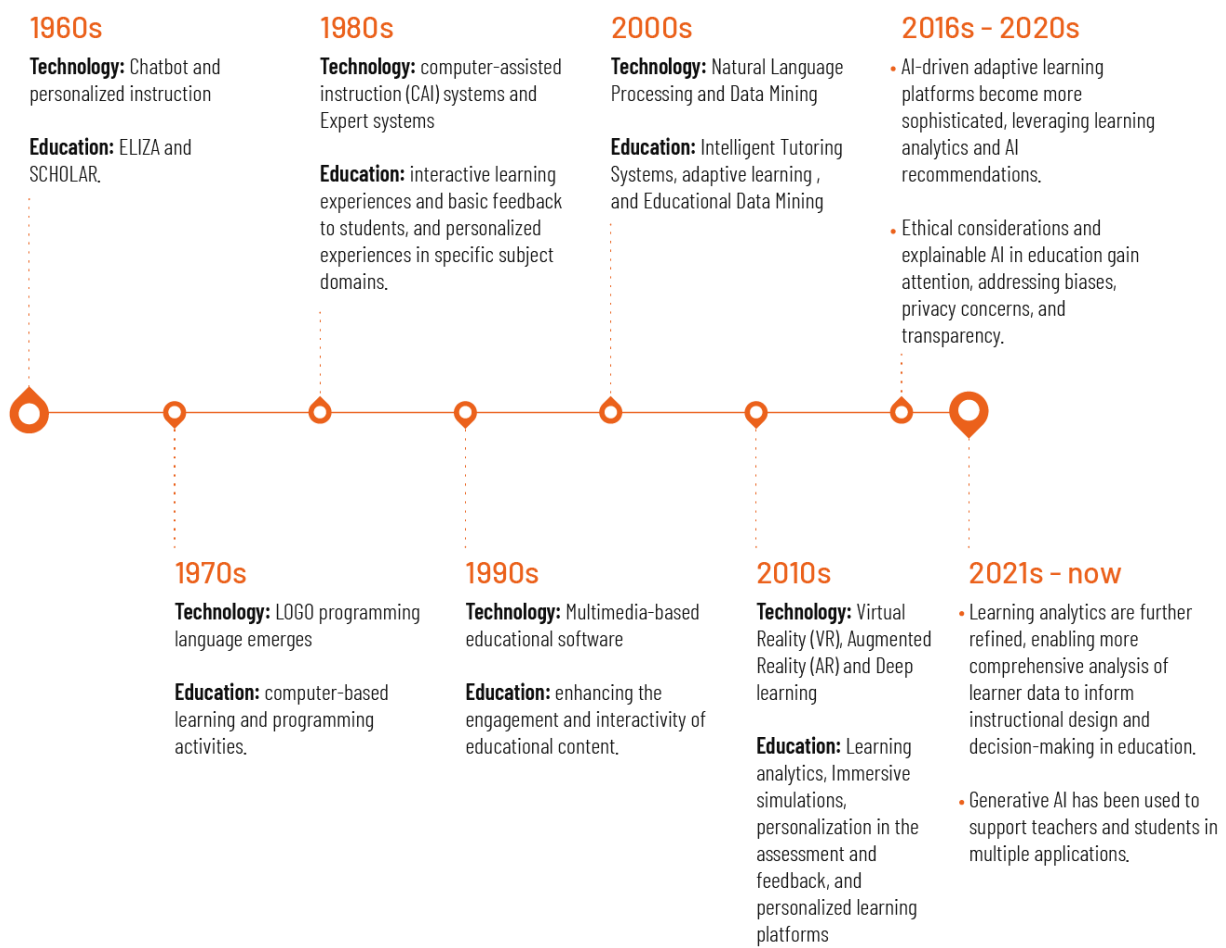

Figure 1: Brief overview of the AIED evolution

This timeline highlights that AIED has been at the center of attention for a long time and encompasses a wide variety of technologies. As in other fields, AI continuously redefines how people (e.g., students and teachers) interact with the world (e.g., physical classrooms, online environments, and new learning methodologies). AI solutions can provide personalization and good-quality feedback for students while supporting teachers in assessment activities to create content for their classes. It can revolutionize how we learn and set the stage for a future where everyone can access a high-quality education.

It is of considerable interest to highlight the rapid infusion of diverse AI technologies into this domain. Techniques such as Natural Language Processing, Deep Learning, and the immersive experiences enabled by Virtual Reality (VR) and Augmented Reality (AR), powering the AI with the huge amount of data, serve as exemplary instances of technologies swiftly integrated into educational paradigms, thus marking a significant shift in pedagogical approaches. Based on current trajectories, we anticipate that Generative AI will emerge as a cornerstone technology in education.

# Generative Artificial Intelligence

Generative Artificial Intelligence (GenAI) harnesses the power of deep learning networks to discern meaningful patterns and relationships within vast databases. By doing so, it can generate original and innovative artifacts across various sources, including images, texts, audio, music, videos, and more [7]. These advanced models hold immense potential to change how we interact with technology, introducing transformative shifts in numerous daily processes. With its ability to create novel outputs, GenAI has the capacity to reshape traditional paradigms and pave the way for innovative approaches in various domains [8].

The advancements in GenAI in the last few years occurred due to various factors, including the enhanced computational capacity, the abundance of data, the continuous refinement of deep learning algorithms, and the substantial investments of big tech companies in the topic. This powerful combination has enabled scientists and researchers to tackle an expanding range of problems characterized by high levels of complexity with remarkable success [7]. Leveraging the vast quantities of training data at its disposal, GenAI has demonstrated an impressive capacity to tackle a broad spectrum of issues, augmenting applications across various sectors. Not only has GenAI achieved remarkable results in tasks such as image synthesis, text generation, and drug discovery, but its rapid evolution is also forging the path toward solutions that were previously thought to be out of reach. From highly personalized content recommendations to creating entirely new designs in architecture and fashion, GenAI has underscored its potential to drive innovation on a grand scale.

The recent popularization of GenAI was due to ChatGPT, a Large Language Model (LLM) [9]. Utilizing a sequence-to-sequence algorithm, LLMs operate under the principle of generating novel content in response to an input text. This means the model requires an initial textual prompt to stimulate the production of entirely new, synthesized content, creating a dynamic and interactive experience.

Although the advances noticed recently, LLMs benefit from concepts and algorithms that go back to the 1940s and 1950s, such as Neural Networks and Word Embeddings. The most recent techniques used are the Transformers networks with Attention [10]. In short, word embedding techniques aim to create numerical representations using vectors to capture the meaning of words. In this vector space, words with similar meanings are typically located closer to one another.

Transformer is a deep learning architecture designed to create sequence-to-sequence models [9]. Its unique ability to consider several critical factors when performing modeling is the main novelty. The Transformer architecture internally considers the order of the text, word disambiguation mechanisms, and increases the importance of key concepts, leading to more context-aware outcomes. This advancement in model comprehension significantly enhances its ability to produce accurate and meaningful results. The initial application of Transformers was related to text translation, but this is the most relevant technical aspect of an LLM.

It is crucial to emphasize that GenAI goes beyond text generation, encompassing the synthetic creation of images, videos, and music, which has recently gained significant attention. While this application is not novel to AI, it has seen remarkable advancements and benefits from traditional methods. For instance, as early as 1957, "The Illiac Suite" composed by Lejaren Hiller and Leonard Isaacson [13] stands as a pioneering example of computer-generated music in electronic music's history.

In this context, Generative Adversarial Networks (GANs) have emerged as a recognized technique [14]. GANs operate on the contest principle, employing two neural networks (a generator and a discriminator) to collaborate in generating synthetic yet highly realistic data. The generator network continually refines its output to deceive the discriminator network, resulting in strikingly realistic synthetic data, whether images, audio, or even video content. This architecture effectively leverages Convolutional Neural Networks (CNNs), which is a deep learning architecture to process visual data. Similar to word embeddings, GANs rely on high-dimensional vector spaces to generate content. For example, a point in this high-dimensional space can serve as input to the GAN, yielding a corresponding image as output. Over time, the GAN learns to map various parts of this high-dimensional space to recognizable objects, creating a 'latent space' of possible outputs. This forms the foundation of how GANs can generate diverse content from seemingly random inputs.

In education, GenAI can create novel content that aligns with learned patterns, changing the dynamics of pedagogical methodologies and the dissemination of knowledge. Its potential applications, from personalized learning content to advanced tutoring systems, represent a fundamental shift towards a more individualized, adaptive, and diverse educational landscape. In the next section we present a few cases of practical adoption of GenAI in Education.

# GenAI in Education

This section illustrates the adoption of GenAI in Education, showcasing cases such as Khan Academy, Coursera, Duolingo, and its widespread integration in leading universities worldwide. These examples exemplify how GenAI can shift educational experiences and enhance learning outcomes. Notably, the prominent EdTech and universities have long been at the forefront of AI implementation in education. Therefore, the adoption of GenAI has been built upon the existing foundations, leading to a fluid and progressive integration. The following sections show traditional uses of AI and the developments that led to GenAI adoption.

# Personalized Learning

Personalized learning has been one of the primary targets in harnessing AI for educational advancements since its beginning. It adopts AI to facilitate individualized learning experiences adapted to each student's needs [15]. For example, Harvard and MIT Universities co-founded OpenEdX, an AI-based platform that offers personalized course recommendations and adaptive learning pathways informed by learner data [16]. By intelligently analyzing learner data, the platform identifies learner strengths and areas of improvement, subsequently offering a curated selection of courses that align with individual interests and aspirations. This personalized course recommendation system enhances student engagement and motivation.

In this light , Coursera, a leading online learning platform, also applies AI to provide a personalized learning experience for students. It benefits from AI algorithms to assess learner progress, identify areas of strength and weakness, and dynamically adjust the learning path. By personalizing the content and activities, Coursera helps students focus on areas that require further attention and accelerates their learning progress [17].

Duolingo, a language-learning platform, applies AI to its adaptive learning methodology [18]. The platform evaluates learners' performance, tailoring the complexity of exercises and lessons per their progress. This customized approach ensures that learners are given the appropriate level of challenge and receive targeted instruction tailored to their learning needs [18].

Finally, Khan Academy utilizes AI algorithms and data analytics to analyze individual student data, including their performance, progress, and learning preferences [19]. With this data, Khan Academy generates personalized learning paths for each student. These learning paths recommend specific content and exercises tailored to students' needs. By adapting to the individual pace and focusing on areas that require more support or challenge, students can learn at their speed and achieve optimal learning outcomes.

Recently, Khan Academy developed the Khamigo [20] AI-powered guide for teachers and students. It uses the power of LLMs, more specifically GPT-4, to do many tasks, such as analyzing student responses and providing instant feedback. Khamigo creates a personalized and adaptive learning experience for students. It complements traditional instruction by providing tailored support, allowing students to learn at their own pace, and helping them develop a deeper understanding of the subject matter.

Finally, GAN architectures have proven to be instrumental for students to achieve personalized learning experiences. A recent paper published in Nature [21] shows how students can significantly enhance their engagement with the course material and each other through direct interactions with relevant historical figures related to the study context. For instance, while exploring physics, students can now have the opportunity to interact virtually with iconic figures like Einstein, or during an art class, they can engage with the likes of the enigmatic Mona Lisa. This innovative approach immerses learners in the subject by bringing historical luminaries to life, fostering a deeper connection with the content and stimulating curiosity and enthusiasm for learning [21].

# Assessment and Feedback

AI has also emerged as a powerful tool to automate grading processes and offer timely feedback to students. For instance, at the University of Georgia researchers employed AI-powered virtual teaching assistants to assist professors with administrative tasks, answer students' questions, and provide personalized feedback on assignments [22].

In this context, Duolingo deploys AI models to parse learners' exercise responses and provide instantaneous feedback [18]. This feedback incorporates error rectification, explanation, and guidance, enabling learners to understand and amend their errors. By capitalizing on AI, Duolingo provides immediate and personalized feedback, empowering learners to solidify their language comprehension and accelerate their progress more efficiently [18].

Coursera and Khan Academy have been using the power of LLMs (Coursera Coach[17] and Khamigo[20]) to analyze student submissions and offer automated feedback, including error correction, guidance, and explanations. By receiving immediate feedback, students can identify areas of improvement, rectify mistakes, and deepen their understanding of the subject matter. Besides, Coursera Coach also tries to make the platform more accessible by including machine translation models to allow students to take coursework in their own language.

Tutoria exemplifies an additional application of AI aiding in the assessment process. This innovative platform harnesses the power of AI to propose pertinent tags for educators dynamically as they evaluate open-ended questions [23]. In doing so, Tutoria bolsters teachers' efficiency, empowering them to manage large-scale courses effectively. Furthermore, the platform incorporates LLMs to autonomously generate feedback based on the teachers' responses, adding another support layer to their teaching toolkit.

Image generation models such as Dall-E and GPT-4 have been employed to support the creation of practical illustrative examples, like X-ray images, for assessment purposes in medical training. This application has demonstrated promising implications due to the substantial diversity of images that can be generated, thereby providing a richer and more comprehensive learning resource for medical trainees [23].

In conclusion, it is crucial to highlight the growing trend of universities encouraging instructors to integrate GenAI into their daily assessment activities. These institutions typically illustrate the potential of GenAI, offer guidelines on its usage, provide a comprehensive list of tools, and advise on assessment design, among other supportive measures, aiming to enhance teaching efficacy and learning outcomes [25][26][27].

## Predictive Analytics

Predictive analytics uses statistical and machine learning techniques to analyze historical educational data to predict future or unknown educational outcomes. In education, predictive analytics could be used for various applications, such as student performance prediction, student dropout prediction, emotional and cognitive engagement detection, growth and development indicators for college students, and at-risk student identification [28]. For instance, Arizona State University leverages predictive analytics via the "eAdvisor" system [29]. This tool identifies students facing academic hurdles, allowing for timely support and intervention.

Considering commercial tools, Khamigo also utilizes AI-powered learning analytics to track student progress and provide valuable insights for teachers. Data analytics algorithms analyze student performance data, generating comprehensive reports highlighting strengths, weaknesses, and areas requiring attention. This information empowers teachers to make data-informed decisions to support student growth [20].

Furthermore, Coursera's AI-driven analytics provide valuable insights for instructors. Teachers can access learner behavior, engagement, and performance data to understand student needs better and adapt their teaching strategies accordingly. This data-driven approach empowers instructors to continuously improve course delivery and optimize the learning experience for students [17].

It is important to remember that predictions are based on probabilities and are not guaranteed to be accurate. Furthermore, ethical considerations such as data privacy and fairness should be considered when using predictive analytics in an educational setting. In this direction, a recent study goes a step beyond, illustrating how ChatGPT, when paired with generated predictions, could pave the way for a prescriptive analytics strategy. This methodology could significantly increase the explainability of predictive analytics and, in this case, reduce the rate of task non-completion among students, thereby promoting better academic achievement and engagement [30].

## Virtual Teaching Assistants

AI-powered intelligent tutoring systems provide personalized guidance and support to students. A case in point is Carnegie Mellon University's "Cognitive Tutor [30]." This system harnesses AI algorithms to modify instructional content dynamically and deliver personalized feedback to students.

Another noteworthy example of this technology is the AI teaching assistant, "Jill Watson," employed by the Georgia Institute of Technology [31]. Jill, developed using IBM's Watson AI technology, was integrated into an online course to manage the high volume of questions posted on the course's discussion forum. The AI was trained on thousands of past questions and answers, enabling her to provide accurate and instant responses to student inquiries. Jill's interactions were so convincing that most students were unaware they were communicating with an AI until the end of the course. This case underscores the vast potential of LLMs in educational settings, providing prompt responses to students and allowing human instructors to focus on complex tasks, thereby improving the overall educational experience.

Recently, Duolingo unveiled a collaboration with OpenAI on the Duolingo Max project [33], which plans to integrate generative AI into its repertoire. This will launch two new features for learners: 'Explain My Answer' and 'Roleplay.' The former provides a chatbot offering students personalized, timely feedback on incorrect responses. The latter introduces an innovative feature where students can engage in unscripted chats in any language with diverse characters within the Duolingo app.

## Smart Campus Solutions

AI and GenAI-driven technologies have the potential to augment both campus-based and online educational experiences substantially. By implementing intelligent scheduling systems, these technologies can streamline class organization and time management, leading to more efficient and less stressful academic environments. In addition, introducing AI-driven campus security solutions can elevate the safety measures of the institutions, ensuring a secure learning environment for all students and staff. For instance, teaching hospitals apply AI-based systems to streamline room scheduling and resource allocation across their campi [34].

LLMs can be used to make education more accessible. As mentioned before, Coursera allows students to take coursework in their own language [17]. Moreover, the company Verbit adopts AI and LLMs to transcribe and translate lectures and seminars online, providing a valuable resource for hearing-impaired students and non-native English speakers [35].

Universities can use LLMs to generate or supplement educational content. Many researchers have been demonstrating that LLMs can be used to create diverse, engaging learning materials for complex subjects, making them more accessible to students [28,36]. In this direction, Coursera applies LLM models incorporated promises to support the creation of engaging content faster and at scale [17].

Finally, the LLM can ensure academic integrity as a proctoring and anti-plagiarism tool during online exams. LLM will be combined with facial recognition algorithms to detect potential cheating or plagiarism, maintaining a secure assessment environment [37].

# Concerns for GenAI in Education

Alongside with the potential benefits, the use of AI in education also raises concerns that must be carefully addressed. We will delve into these concerns, including equity issues, data privacy, bias, human interaction, and ethical considerations [6]. One significant concern revolves around access to students' data. Data privacy and security are also prominent concerns in using AIED. The collection and analysis of vast amounts of student data raise questions about the security and privacy of personal information. Safeguarding student privacy and ensuring robust data security measures are critical to prevent unauthorized access or misuse of sensitive data [38].

Another concern involves the potential for bias and discrimination in AI systems. AI algorithms are trained on existing data, and if that data contains biases, the AI systems can perpetuate and amplify these biases [39]. In other words, the AI system can produce discriminatory outcomes, such as biased grading or course recommendations. Ongoing efforts are necessary to train AI models responsibly and regularly monitor them to mitigate bias and ensure fairness. In a similar direction, issues related to equity and accessibility of AI services have been discussed by the community. The reliance on AI-powered tools requires access to technology and high-speed internet, which can create a digital divide and limit educational opportunities for students from disadvantaged backgrounds, potentially widening existing inequalities [40].

The lack of human interaction is also a worry associated with employing AI in education [6]. While AI can provide valuable support, the absence of direct human connection and personalized guidance may hinder the development of social and emotional skills in students. For instance, there is a concern that an overreliance on AI may devalue human autonomy and creativity [41]. While AI can provide support, it should not replace the role of educators or undermine the importance of human interaction and critical thinking skills in education. Moreover, it is essential that our attention converges on nurturing students' soft skills such as critical thinking and collaboration, a task achievable through an optimally balanced mix of AI tools and human mentorship. Striking a balance between AI-driven systems and human interaction is crucial for a well-rounded educational experience.

Ethical considerations are of utmost importance in AI-powered education. These considerations encompass crucial elements like the explainability of models, securing informed consent, and ensuring responsible utilization of AI-generated content and assessments [42]. Typically, Gen AI models function as a 'black box,' providing little scope for teachers to interpret the results directly. Consequently, enhancing explainability emerges as a vital strategy to provide useful information for teachers, mitigating this challenge. Transparent, explainable, and ethical AI practices are necessary to maintain trust and uphold the integrity of the educational process.

Finally, the potential pitfalls of over-dependence on AI are manifold. They range from intellectual property and copyright issues to the possible job market transformation, stoking fears of AI replacing human jobs. Moreover, there's the looming specter of escalating psychological problems due to social isolation, among other repercussions. Hence, addressing these concerns requires collaboration among policymakers, educators, technology developers, and other stakeholders. Safeguarding student privacy, promoting equal access, ensuring model explainability, and fostering ethical AI practices are essential steps in harnessing the potential of AI in education while mitigating associated worries. As AI advances in education, ongoing monitoring and proactive efforts to address these concerns will be vital in creating an inclusive, equitable, and ethically responsible AI-powered education system.

# Final Considerations

We conclude this article by stating that incorporating artificial intelligence (AI) into education can redefine our teaching and learning paradigms. By leveraging personalized learning, intelligent tutoring, automated grading, and individualized feedback (among other) AI-powered tools, education could transform into a more adaptive, efficient, and immersive experience. However, merely adopting AI, GenAI, or other technological advancements in educational settings is insufficient to surmount the challenges faced by institutions, teachers, and students. Technological progress must be coupled with the scrutiny and enhancement of learning methodologies, always keeping the stakeholders at the heart of this evolution.

In this direction, it is important to highlight the role of teachers in the AI ecosystem. Providing ample training and support for educators is a critical prerequisite for effectively assimilating AI technologies into their pedagogical practices. By equipping teachers with the requisite training and resources, they can proficiently harness AI tools, interpret AI-generated data, and adeptly maintain an equilibrium between AI-guided instruction and human interaction. It is crucial to reinforce the logic of co-creation, where teachers and AI work together, rather than subscribing to a replacement ideology, where AI supersedes teachers. AI should be viewed as a tool rather than a replacement. Combining human insight and AI capabilities can create a robust, effective, and inclusive educational environment. This co-creation approach not only leverages the advantages of AI but also values the irreplaceable human touch in the learning process. As mentioned before, it is about striking a balance where AI enhances the educational process and teachers provide the emotional intelligence and critical thinking skills necessary for holistic student development.

Continued evaluation and improvement of AI systems in education are crucial to identify strengths, weaknesses, and areas for improvement. Furthermore, the assessment and enhancement of AI systems in practices (i.e., in educational institutions) are indispensable to discern their strengths, shortcomings, and potential areas of improvement. Instituting regular validation processes, feedback mechanisms, and cycles of iterative improvements can help to fine-tune AI applications, thereby ensuring their efficacy in enriching educational experiences. Moreover, proactive strides should be taken to reduce the digital divide and guarantee fair access to AI-empowered educational tools for all. This encompasses addressing biases embedded within AI algorithms, customizing AI applications to cater to diverse learning requirements, and facilitating access to technology and high-speed internet within underserved communities.

In our view, a series of measures are needed to ensure AI's ethical and efficient implementation in education. Persistent research and development efforts are required to advance AI applications in this domain. This includes honing the accuracy of algorithms, augmenting adaptive learning systems, and refining natural language processing techniques to bolster personalized feedback and tutoring. Furthermore, fostering collaboration between educators, researchers, policymakers, and technology developers is a key necessity to develop relevant tools for supporting educational settings for a wider audience. Such a synergistic approach will help identify and propagate best practices, disseminate insights, and formulate comprehensive guidelines and standards for the ethical deployment of AI in educational contexts.

By embracing these next steps, we can harness the full potential of AI in education while mitigating risks and addressing concerns. The use of AI may help to enhance education, empower learners, and support educators in delivering personalized and impactful learning experiences. Adopting a thoughtful and inclusive approach, leveraging AI as a powerful tool to complement human expertise, and fostering a future-ready education system will prepare learners for the challenges and opportunities of the future.

As a final message in this article, we believe a joint effort from different stakeholders could fully unlock AI's potential in education while concurrently mitigating associated risks and addressing pertinent concerns. AI bears the potential to significantly elevate the quality of education, empower learners, and provide robust support to educators in curating personalized and meaningful learning experiences. The following articles of the series Education in the Age of Generative AI features by CESAR will give details on the potential, ethical considerations, and the Brazilian scenario of using AI and Generative AI in education.

# About the Author(s)

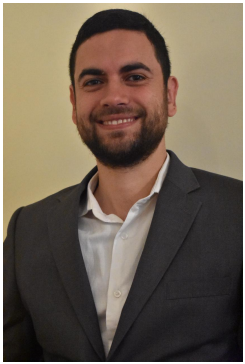

**Rafael Ferreira M.ello** has a Ph.D. in computer science with research interests that span across learning analytics and natural language processing. He is a Professor at the CESAR School in Brazil, where he works with the development of Artificial Intelligence Technology For Education. In the last few years, Dr. Mello has supported the adoption of Learning Analytics techniques and tools in several institutions in Brazil. Dr. Mello worked on several multinational research projects, which involved institutional and organizational partners in Europe, Australia, and Latin America. He has published in the leading international journals and conference proceedings in the fields of his research. He also served as an assistant editor and reviewer for several journals and conferences, including the International Learning Analytics & Knowledge Conference, Journal of Learning Analytics, Computers and Education, British Journal of Educational Technology, and Internet and Higher Education.

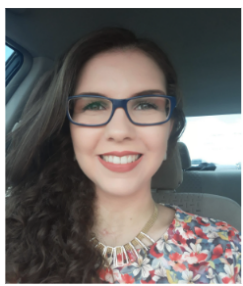

**Elyda Freitas** has a Ph.D. in Computer Science. She is a Professor at the University of Pernambuco and post-doc researcher at CESAR School, where she is involved in projects regarding the adoption of Learning Analytics. She has published her results in national and international journals and conferences, such as Revista Latinoamericana de Tecnología Educativa, Brazilian Journal of Computers in Education, and Brazilian Symposium on Informatics in Education - having worked as a reviewer in these last two.

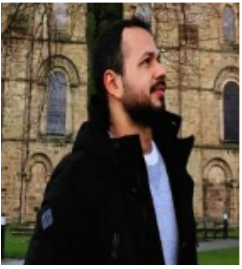
**Filipe Dwan Pereira** received the B.S. degree in Computer Science from Federal University of Roraima and the doctorate in Artificial Intelligence Applied to Education from Federal University of Amazonas. Currently he is conducting a post-doc research on natural language processing (NLP) at CESAR School Brazil. Since 2013, he has been an Assistant Professor with the Department of Computer Science, Federal University of Roraima. Dr. Pereira collaborated on several multinational research projects involving institutional and organizational partners in Europe, New Zealand, and the USA. His area of research covers education data mining, learning analytics, artificial intelligence, machine learning, big data, computing in education, NLP, and information systems.

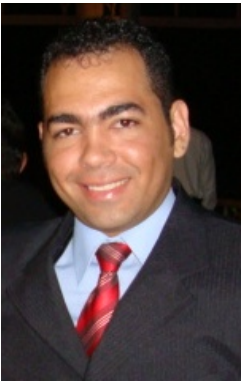
**Luciano de Souza Cabral** has a Ph.D. in Electrical Engineering with research interest in Artificial Intelligence using Natural Language Processing, Machine and Deep Learning applied to Education, Health among others fields. He is professor at IFPE and post-doc researcher at C.E.S.A.R School, where he works with the development of AI Tech for Education. He has worked in research and R&D projects involving AI applied for Education for Brazilians and International organizations. He has published papers on national and international journals and conferences, books, patents, software, among others technical productions. He has participated as reviewer for several journals and conferences including Springer Nature Computer Science, Applied Soft Computing, Computers in Human Behavior, BRACIS, among others.

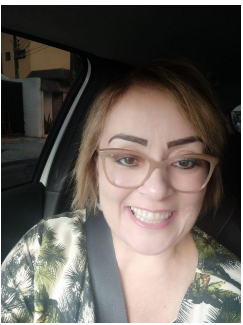
**Patricia Tedesco** is a Bachelor in Computer Science, graduated from the Federal University of Pernambuco in 1994. She obtained her MSc from the same university in the area of Artificial Intelligence in Education in 1997 and completed her PhD in the Computer Based Learning Unit - University of Leeds in 2001. She has been a professor in Artificial Intelligence in the Center for Informatics at UFPE since 2002. She has since then participated in many program committees and supervised many MSc and PhD works in the areas of AIED and Distance Learning.

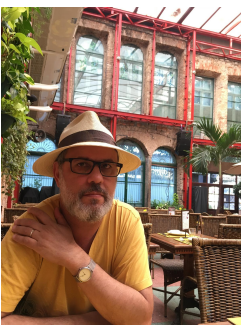
**Geber Ramalho**, 57, electronic engineer, holds a Ph.D. in Artificial Intelligence (AI) from the University of Paris VI in 1997. He is a professor at the Center for Informatics at UFPE, where he conducts research in the fields of AI, interactivity, and innovation. He spent a sabbatical year as an invited researcher at CNRS (France) in 2005 working in the field of generative AI. He coordinated several national and international research and innovation projects. He was a member of the board of the Brazilian Computer Society, and he is currently a board member of CESAR and Porto Digital. He teaches the course 'Ethics in Artificial Intelligence.' He collaborated with the Ministry of Foreign Affairs in the group of experts on Ethics and Technology issues at the United Nations.